# Laser-driven ion acceleration in long-lived optically shaped gaseous targets enhanced by magnetic vortices

I. Tazes[1,2], S. Passalidis[3,4], G. Andrianaki[1], A. Skoulakis[1,2], C. Karvounis[1,2], D. Mancelli[1,2,5], J. Pasley[6], E. Kaselouris[1], I. Fitilis[1,2], M. Bakarezos[1], E.P. Benis[7], N. A. Papadogiannis[1], V. Dimitriou[1] and M. Tatarakis[1,2*].

[1]Institute of Plasma Physics and Lasers-IPPL, University Research and Innovation Centre, Hellenic Mediterranean University, 74100 Rethymno, Greece.
[2]Department of Electronic Engineering, Hellenic Mediterranean University, 73133 Chania, Greece.
[3]CEA, DAM, DIF, F-91297 Arpajon, France.
[4]Université Paris-Saclay, CEA, LMCE, F-91680 Bruyères-le-Châtel, France.
[5]University of Bordeaux, CNRS, CEA, CELIA, Talence F-33400, France
[6]York Plasma Institute, School of Physics, Engineering and Technology, University of York, York, YO10 5DD, United Kingdom
[7]Department of Physics, University of Ioannina, 45110 Ioannina, Greece.

*Corresponding author: mictat@hmu.gr

This research demonstrates high-repetition-rate laser-accelerated ion beams via dual, intersecting, counterpropagating laser-driven blast waves to precisely shape underdense gas into long-lived near-critical density targets. The collision of the shock fronts compresses the gas and forms steep density gradients with scale lengths of a few tens of microns. The compressed target persists for several nanoseconds, eliminating laser synchronization constraints. Measurements of multi-MeV ion energy spectra are reported. 3D hydrodynamic simulations are used to optimize the density profile and assess the influence of the Amplified Spontaneous Emission of the femtosecond accelerating laser pulse. A synthetic optical probing model is applied to directly compare simulations with experimental data. 3D Particle-In-Cell simulations reveal the formation of multi-kT, azimuthal magnetic fields, indicating Magnetic Vortex Acceleration as the main acceleration mechanism.

Laser-driven ion sources are of current interest in a range of applications. Ion-driven fast ignition of inertially-confined fusion fuel [1,2] offers several advantages relative to electron-driven fast-ignition schemes. Proton-boron fusion experiments [3,4] in the pitcher–catcher configuration [5,6] are also reliant on our ability to field bright ion-sources. Laser-produced ion sources are also being considered for possible application in tumor treatment [7,8]. There are also many current examples of their use as radiographic sources in high energy-density physics experiments [9,10]. Most breakthroughs have been achieved with over-dense targets by using the Target Normal Sheath Acceleration (TNSA) [11–13] or Radiation Pressure Acceleration (RPA) [14–16] mechanisms. Proton energies up to 150 MeV have recently been reported using these foil-based approaches [17–19] However, conventional solid targets present several limitations. Targets are destroyed after each shot and require continuous replacement, repositioning, and alignment, restricting their suitability for high-repetition-rate ion sources. Dense plasmas produced from solid targets reflect a large portion of the laser energy, resulting in limited coupling through hot electrons [20]. Additionally, solid targets generate debris that can damage optical components. Recent progress in target fabrication has led to designs incorporating foams [21], nanowires and nanostructures [22,23], or multilayered foils [24] These modifications improved laser energy-coupling to the plasma and enhanced proton acceleration efficiency. Efforts have also focused on fast, motorized target delivery systems, such as rotating solid tape setups, to address repetition rate limitations [25,26].

Ion acceleration mechanisms in the near-critical density regime offer favorable energy scaling with laser power. The regenerative nature of these targets enables high-repetition-rate operation. The targetry is approached through various methods, including cluster gas-jets [27,28], cryogenic targets [29,30], liquid droplets [31], and exploding foils [32], where the main acceleration mechanisms include Collisionless-Shock Acceleration (CSA) [33–35] and Coulomb Explosion [27,36]. Efforts have also been made on optical shaping of underdense target density profiles [37–41]; however, challenges in generating long-lived, controlled and reproducible near-critical density profiles have significantly limited the effectiveness of this approach.

In the past two decades, ions have been accelerated to multi-keV and multi-MeV energies from near-critical density targets in experiments using the CSA mechanism [34,37,42,43]. These results were obtained with high-energy, ns or ps, long-wavelength laser systems. Such systems operate at lower intensities and are limited to single-shot operation but reach the near-critical density regime at lower densities due to their longer wavelengths. Recently, 80 MeV protons were measured from the interaction of a near-critical density gaseous target with a petawatt Ti:Sapphire (Ti:Sa) laser [44]. Particle-In-Cell (PIC) simulations predict proton acceleration to energies from hundreds of MeV up to the GeV range via the Magnetic Vortex Acceleration (MVA) mechanism, driven by intense, femtosecond TW–PW laser pulses [45–49]. MVA [45,48–51] is widely considered to be the most promising acceleration technique in terms of scaling with laser power, but experimental evidence of ion acceleration through MVA remains scarce [52–54], particularly for Ti:Sa laser wavelengths. Simulations indicate that precise control of target density, density-gradient and thickness are required [49] for this acceleration mechanism to be practicable.

This letter reports the first experimental demonstration of high repetition rate ion acceleration using a newly developed novel method for optically shaping gas-jet targets. In this concept, the initial underdense gas-jet profile is shaped using two blast waves generated by two ns duration laser pulses intersecting at 60° in the gas-target. This angle is determined to be the optimal, based on a systematic parametric hydrodynamic simulation study. This approach enables maintenance of the compressed gas profile for more than 15 ns, resulting in a broad synchronization window between the shaping and main acceleration pulses. Additionally, it eliminates unwanted long-scale-length pre-plasma, enabling undisturbed propagation of the main pulse to the steep density gradient. Parametric computational studies indicate a larger compression factor compared to planar colliding blast wave geometries under similar laser and target conditions [55,56]. The observations are supported by 3D hydrodynamic and PIC simulations, which reveal the underlying role of the induced magnetic field vortices in the acceleration mechanism. Accelerated ion spectra with cut-off energies exceeding 11 MeV/u are observed, featuring a quasi-monoenergetic peak.

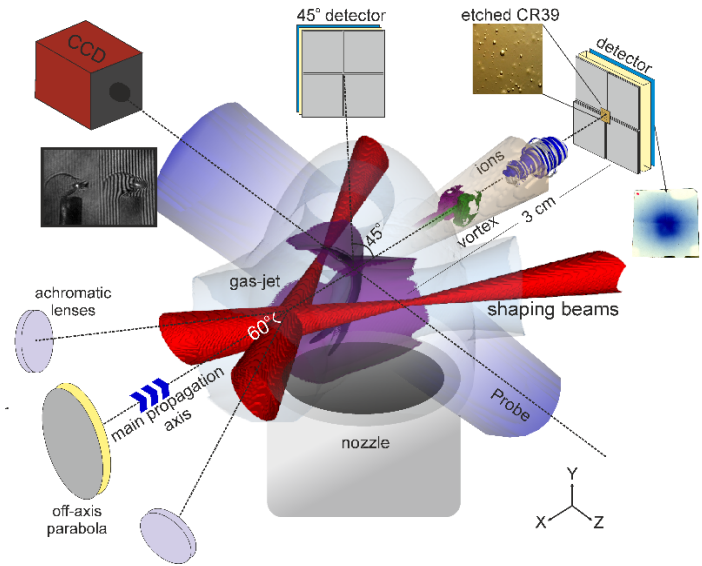

FIG. 1. Schematic layout of the experimental setup. The shaping beams drive dual intersecting blast waves, compressing the gas-jet profile. The black onyx-shaped region indicates peak compression. The delayed main pulse propagates undisturbed through a hollow pathway, penetrating the compressed target and accelerating multi-MeV per nucleon ions. Ion acceleration is supported by the generation of a kT-scale magnetic vortex field. A probe beam records the evolution of the compression on the ns scale and the accelerated plasma structure on the ps scale. Shadowgraphy and interferometry are acquired simultaneously using a Nomarski diagnostic. Passive detectors, CR39 and radiochromic film record the ion and electron signals.

The experiment was performed using the Zeus 45 TW Ti:Sa laser system located at the Institute of Plasma Physics and Lasers (IPPL) [57]. The experimental realization of the conceptional background is illustrated in Fig. 1. An 800 nm, 25 fs FWHM, p-polarized laser pulse with 1 J energy was focused in a 3 μm diameter spot, using an f/2 parabolic mirror, resulting in a normalized vector potential amplitude $a_o$~14. An 800 μm diameter cylindrical nozzle delivered an underdense gas-jet composed of a 99% He–1% $H_2$ mixture at 55 bar. The measured density was $9 \times 10^{19}$ /cm$^3$ corresponding to 0.05 $n_{cr}$. A 1064 nm, 6 ns FWHM laser pulse with 0.85 J energy was split into two equal parts and focused on both sides of the gas-jet forming two ~15 μm spots to generate the dual blast waves. Furthermore, a 10 mJ, 800 nm probe pulse was used to diagnose the interaction with ~fs timing accuracy, controlled by a motorized delay stage with 30 nm step size. The probe pulses passed through a Wollaston prism generating two beams of different polarization, which partially interfere, resulting in Nomarski interferometry along with shadowgraphy, captured simultaneously by the same CCD camera. As particle diagnostics, the well-established CR39 nuclear track detectors covered by a multilayer aluminum filter mask were used, allowing measurement of different ion stopping powers per mask sector. The CR39 detector was placed 3 cm from Target Chamber Center (TCC). The dose distribution of ions and electrons was acquired using a radiochromic film (RCF), placed behind the CR39.

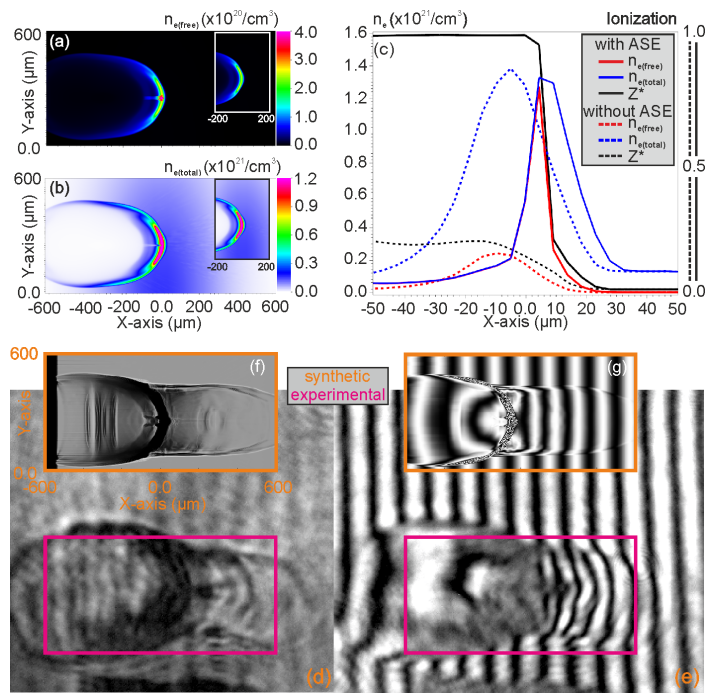

FIG. 2. FLASH simulation results of free electron density (a) and total electron density (b) at maximum compression, with and without ASE effects (insets), 5 ns after the arrival of the leading edge of the shaping pulses at the target. Lineouts of electron densities and ionization degree Z* (c), with ASE (solid lines) and without ASE (dashed lines), show the steepening of the density profile. Comparison between experimental shadowgraph (d) and interferogram (e) with synthetic ones (f and g) at similar times illustrates an overall good agreement. To capture ASE effects, the laser attenuator was set to 0%, allowing only a few mJ of the main pulse energy to reach the target. Interferograms show qualitative agreement in fringe shifts. A larger region is obscured by probe absorption.

Additionally, 3D FLASH [58] hydrodynamic simulations were performed to study the temporal evolution of the shaping. The simulations included the effects of the inherent Amplified Spontaneous Emission (ASE) of the main accelerating pulse. A synthetic diagnostic model has been developed to perform ray tracing within the simulation domain, producing synthetic interferograms and shadowgraphs. This approach enables direct comparison between the measurements and the simulation results, rather than the optical probing results with simulated density profiles. The synthetic diagnostic model accounts for the index of refraction from both electron and neutral species, as well as absorption of the rays [59–61].

Details of the dynamics of gas-shaping ion acceleration conditions are presented in Fig.2. The ns shaping pulses fully ionize the gas at the focal spot region while the shock fronts of the generated blast waves remain predominantly neutral [Fig. 2(c), dashed lines] [62]. As a result, the probe pulse encounters the index of refraction associated with the ejected electrons [Fig. 2(a) and (c) red lines]. The neutral density of the shock fronts contributes negligibly to the refractive index [34], and therefore, shadowgraphs and interferograms are the result of gradients in the ionized electron density, $n_{e(free)}$, and associated index of refraction. When the fs main accelerating pulse arrives, it interacts with the shaped near-critical neutral gas density profile ejecting two electrons per helium atom, resulting in the electron density, $n_{e(total)}$ shown in Fig. 2(b) and (c) [blue lines]. The prepulse preceding the main interaction ionizes the neutral shock front [Fig. 2(c), solid lines] and thus, the main fs pulse interacts with a near-critical electron density formed by the prepulse.

A significant advantage of the present gas-jet shaping scheme is that even though intense ASE (intensity contrast ~ $10^6$) can be destructive for thin foil targets [63,64], here it contributes beneficially by steepening the density profile. This can be clearly seen in Fig. 2(a, b and c) where the profile thickness shifts from 40 μm at 1/e of the peak density without ASE to 23 μm with ASE, with the peak density remaining almost unchanged. The lack of cylindrical symmetry prevents reliable Abel inversion for cubic electron density retrieval. Therefore, direct comparison between synthetic diagnostics and experimental probing is required. Experimental and synthetic shadowgraphs [Fig. 2(d) and (f)] and interferograms [Fig. 2(e) and (g)] show good overall agreement, which suggests consistency between experimental and numerical results for both electron and neutral densities, thus validating the numerical model. The total electron density reaches ~1.2 × $10^{21}$/cm$^3$ (~0.69 $n_{cr}$), which according to the MVA matching condition [48,49], corresponds to an optimum channel length of ~35 μm for the specifications of the fs main accelerating pulse.

During the ion acceleration phase, the shaped target is irradiated by the 45 TW fs laser pulse. The results of the averaged signal from 50 shots, accumulated on the passive diagnostic detectors, are presented in Fig. 3. The red arrow in Fig. 3(a) indicates the propagation direction of the main pulse. The interaction shows intense point-like self-emission at the laser–shock front interface [Fig. 3(a) and (b)], which is a characteristic feature of intense laser: critical-density interactions [43]. The self-emission remains intense despite the presence of neutral density filters and an 800 nm interference filter. Optical probing diagnostics show that the pulse penetrates the near-critical density region, forming an accelerated plasma structure at the rear of the target, as captured at 2 ps after the interaction [Fig. 3]. The approach is insensitive to the jitter between the synchronized laser systems (±2 ns in our experiments). Fig. 3(a)–(f) shows shadowgraphs and interferograms of the interaction between the main pulse and the compressed target at 5, 9, and 13 ns, from three separate shots (time zero corresponds to the arrival of the start of the ns shaping pulses at the target). The acceleration interaction remains consistent throughout the compression duration of approximately 15 ns, with optimal synchronization at ~5 ns, as indicated by hydrodynamic simulations. No plasma channel is observed on the left side of the shaped target, both experimentally and in simulations, indicating a hollow pathway for undisturbed propagation of the main pulse. This is crucial for the unperturbed propagation of the intense main fs laser pulse.

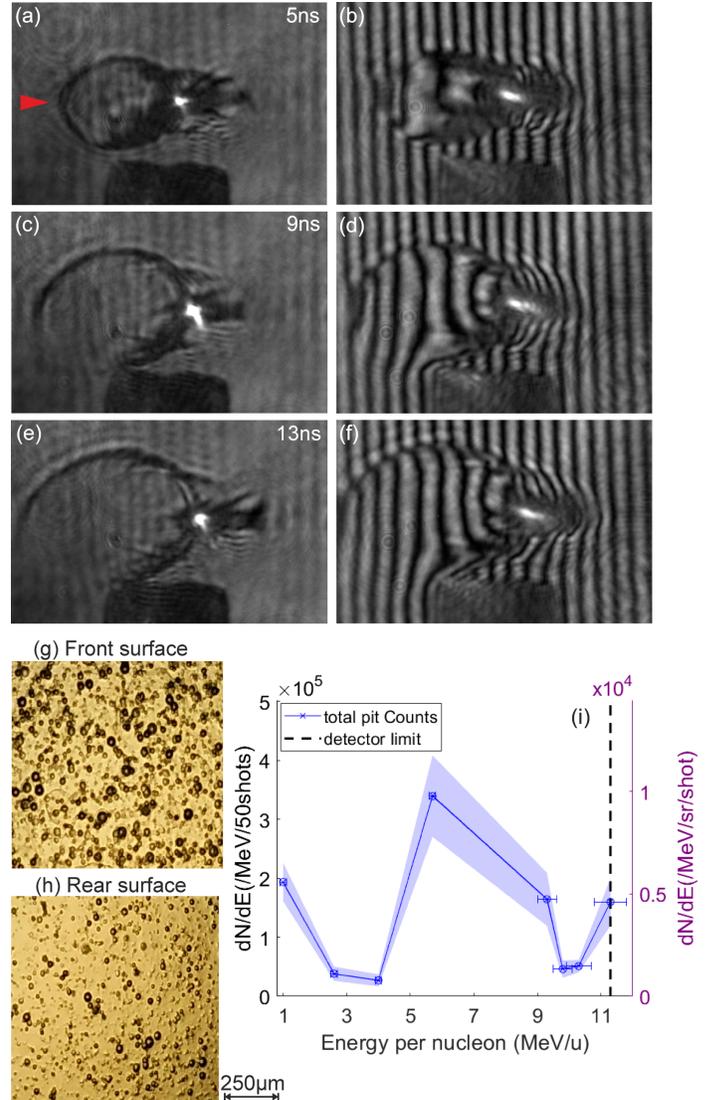

FIG. 3. Shadowgraphs (a), (c), (e) and interferograms (b), (d), (f) from three separate shots, showing the evolution of the compression at 5, 9, and 13 ns, respectively. The main interaction with the target is observed at 2 ps in all cases. The red arrow indicates the propagation axis of the main pulse. Front (g) and rear (h) surfaces of the etched CR39 under transmissive microscopy show pits generated by accelerated ions, at 5.7 MeV/u and 11.3 MeV/u energy diagnostic sectors, respectively. The ion energy spectrum (i) is extracted from CR39 by measuring pits corresponding to different Bragg peaks.

The multilayer filter mask of the CR39 detector is segmented into eight sectors. The four front-surface sectors correspond to energies ranging from 1.0 to 5.7 MeV/u, while the four rear sectors cover the 9.3 to 11.3 MeV/u range. Ion stopping power is calculated using SRIM Monte Carlo simulations [65] to calibrate the detector. The lowest energy detection threshold of CR39 is 100 keV, determined using an uncovered region, not exposed to the IR laser pulse, preventing laser-induced damage. After chemical etching in NaOH solution, pits are counted under transmissive microscopy to extract the ion energy spectra [Fig. 3(g) and (h)].

The energy spectrum demonstrates a quasi-monoenergetic peak feature around 6 MeV/u and the cut-off ion energy at 11.3 MeV/u. Fewer large pits, associated with higher-Z ions, are observed on the rear side of the CR39. This indicates a greater number of higher energy-per-nucleon protons due to higher charge-to-mass ratios, reaching the rear surface. Integrating along the energy spectrum, a total number of $4.2 \times 10^6$ ions/shot is obtained. 3D PIC simulations predict ~10 MeV/u protons and ~5 MeV/u helium ions (20 MeV total energy) [Fig. 4(d)]. A Gaussian-like, high-dose distribution of ~1.1 Gy per shot is recorded on the RCF, from both ion and electron signals. As expected, no measurable ion signal appears on the CR39 detectors when one or both of the ns shaping laser beams are switched off. In order to investigate the underlying acceleration mechanism in the current gas-jet shaping conditions, EPOCH [66] 3D PIC simulations were employed [55,56]. The PIC models use initial conditions derived from hydrodynamic simulations, consistent with experimental observations. The initial electron density, with a peak of $1.2 \times 10^{21}/cm^3$, responds rapidly to the propagation of the focusing main fs laser pulse. This interaction generates plasma wakes, forming a short-scale-length LWFA structure [Fig. 4(a) and (b)]. Beyond the peak density, at 300 fs, a large electron population is injected into the bubble, producing a high current of energetic electrons (>100 MeV). The bubble elongates significantly after the electron density peak, a process that appears to facilitate electron injection, similar to the down-ramp injection mechanism [67,68]. The pulse undergoes relativistic self-focusing just before reaching the peak electron density [Fig. 4(e)] and is efficiently absorbed by the dense plasma [55]. The high electron current drives a strong azimuthal magnetic vortex [Fig. 4 (f)] with a peak value of 100 kT at the time of reaching the peak density region, fast expanding within the plasma waveguide [45,48,50,51]. The transient magnetic vortex induces longitudinal quasi-static electric fields, which along with charge separation, accelerate the protons and helium ions. Concurrently, the magnetic field pinches the proton and ion densities towards the propagation axis of the pulse. The longitudinal electric field is of the order of the electric field of the laser pulse [46]. This strongly indicates that MVA is the dominant underlying ion acceleration mechanism [46-50]. The protons and ions are accelerated in the forward direction in a half-angle of 30° [Fig. 4 (g)] and 20° [Fig. 4 (h)], respectively. Our EPOCH 3D PIC simulations with the laser pulse reaching the peak density of the shaped target, generating the vortex field, and collimating the ion density [Fig. 4 (c)], bear close resemblance to the experimental results. As a result, our study strongly suggests that in the experiments a strong ~kT magnetic vortex drives the acceleration of the ions.

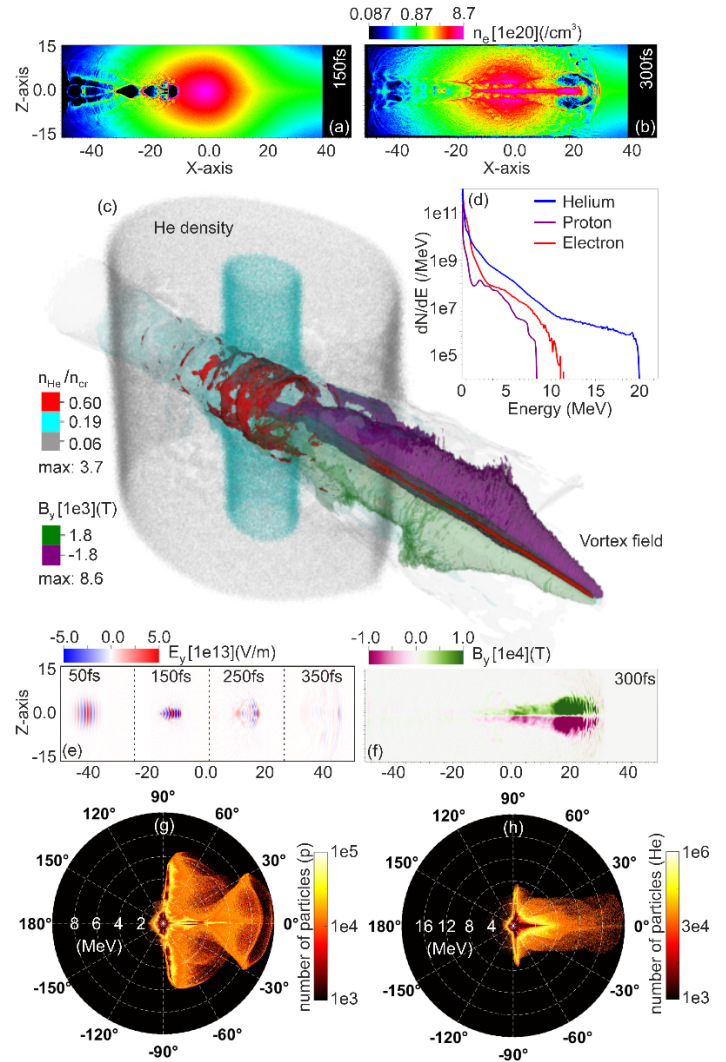

FIG. 4 Electron density at 150 fs (a) and 300 fs (b) from PIC simulations, showing the generation of a short-scale wakefield. 3D iso-surface plot of helium density (in grey, cyan and red) with magnetic fields (in purple and green) at 600 fs (c), illustrating near-critical target–laser interaction. Particle energy spectra (d) of protons, electrons, and helium ions at 850 fs. Longitudinal electric field of the laser pulse between 50 and 350 fs (e), demonstrating strong self-focusing and beam absorption. Y-component of the magnetic field (f) which shows the generation of an azimuthal vortex magnetic field. Proton (g) and helium ion (h) angular energy distributions, demonstrating particle beam directionality with 30° and 20° half-angles, respectively.

In conclusion, we present an exciting demonstration of the production of high repetition rate ion beams from laser-driven MVA. Efficient compression of underdense gas from an 800 μm diameter cylindrical nozzle, into near-critical, sharp-gradient, density profiles, is employed. The required density profiles persist for ~15 ns, which is substantially longer than can be achieved via other means. This eliminates laser synchronization challenges. Our study advances the field towards high-repetition rate ion acceleration applications, eliminating the drawbacks of single-shot and shots-per minute methods, present in high energy low-rep rate laser systems and solid-target approaches. The experiments were conducted with approximately 10-second intervals to allow chamber vacuum recovery, enabling a ~0.1 Hz repetition rate. A record ion energy, exceeding 11 MeV/u, using gaseous targets is reported for high rep-rate multi-TW laser systems. The method is adaptable for use on laser facilities with powers from TW up to PW and for both low and high-repetition rate systems. Work is underway to implement differential pumping and pellicle shields to protect compressor gratings, targeting operation at ~Hz frequencies.

Near-critical density gaseous targets are challenging to characterize. Accurate measurements are required to optimize shaping and acceleration mechanisms. For this purpose, a synthetic diagnostic ray-tracing model has been developed to match optical probing results with simulations, providing indirect measurements that replicate the experimental density profiles. In addition, electron deflectometry is considered the appropriate method to measure high-resolution density profiles along with the magnetic field topology.

**Acknowledgments**
This work is supported by the 'Regional Development Program of Crete' in the context of the project with title 'Use of secondary plasma radiation sources generated by high-power laser–matter interaction for advanced biomedical applications' KA 2024ΝΠ10200000 MIS 5220198.
We acknowledge funding by the Hellenic Mediterranean University within the project "Proposal for post-doctoral research at the Institute of Plasma Physics and Lasers (IPPL) of Hellenic Mediterranean University (HMU)" on the context of the 2607/Φ120/04-05-2022 call of HMU for post-doctoral research.
We acknowledge the support with computational time granted by the Greek Research & Technology Network (GRNET) in the National HPC facility ARIS-under project ID pr016025-LaMPIOS III.